\documentclass[a4paper,11pt]{article}
\usepackage{pos}

\title{ALARIC: A NLL accurate Parton Shower algorithm}
\author*[a]{Florian Herren}

\affiliation[a]{Physik-Institut, Universit\"at Z\"urich,\\ Winterthurerstrasse 190,
CH-8057 Z\"urich, Switzerland}

\emailAdd{florian.herren@physik.uzh.ch}

\abstract{In this contribution, we summarize the key concepts behind the next-to-leading-logarithm accurate parton shower algorithm ALARIC. We discuss tests of the logarithmic accuracy and present comparisons to LEP data.}

\FullConference{Loops and Legs in Quantum Field Theory (LL2024)\\
 14-19, April, 2024\\
Wittenberg, Germany\\}

\begin{document}
\maketitle

\section{Introduction}
Parton shower algorithms are ubiquitous in Monte Carlo event generators for collider experiments. They evolve high energy partons, quarks and gluons, down to the hadronization scale through subsequent soft and collinear splittings. These splittings correspond to the logarithmically enhanced regions of QCD matrix elements and thus parton showers resum the corresponding logarithms. A priori, parton shower algorithms are only accurate at the leading logarithmic (LL) level for generic observables and possibly accurate at the next-to-leading-logarithmic (NLL) level for a select few observables \cite{Dasgupta:2018nvj}. Yet, to better assess parton shower uncertainties and improve event generation for the LHC experiments, parton showers at NLL or above are required. To this end, a significant amount of work has been dedicated to understanding the relevant features of parton shower algorithms critical to the logarithmic accuracy and first NLL accurate showers such as the PanScales showers \cite{Dasgupta:2020fwr} have been developed.

Despite this progress, no publicly available NLL accurate shower has been implemented in a general purpose event generator and compared to actual data. To this end, we developed the ALARIC algorithm \cite{Herren:2022jej} and found an analytic proof that it is NLL accurate, while also maintaining a simple kinematics mapping, facilitating the matching to fixed-order calculations.

Here, we present the key concepts, show first results and summarize further developments.

\section{Key concepts}
In the following we discuss the two main concepts relevant to the NLL accuracy of ALARIC: the treatment of soft radiation and the recoil scheme. In addition, we briefly describe the matching to next-to-leading order (NLO) calculations.

\paragraph{Soft radiation}
In the soft limit the squared matrix element factorizes as
\begin{equation}
    _{n}\langle1,\ldots,n|1,\ldots,n\rangle_{n}=-8\pi\alpha_s\sum_{i,k\neq j}
    \,_{n-1}\big<1,\ldots,j\!\!\!\backslash,\ldots,n\big|{\bf T}_i{\bf T}_k\,w_{ik,j}
    \big|1,\ldots,j\!\!\!\backslash,\ldots,n\big>_{n-1}\;,
\end{equation}
where $j$ is the label of the soft gluon, $i,j$ denote the constituents of the dipoles, the ${\bf T}_i$ are colour insertion operators and
\begin{equation}
    \label{eq:soft_eikonal_intro}
    w_{ik,j}=\frac{p_ip_k}{(p_ip_j)(p_jp_k)}\;,
\end{equation}
is the Eikonal factor. To avoid double counting of soft-collinear contributions, the Eikonal factor has to be distributed properly over the partons $i$ and $k$, such that its collinear limit can be matched to the collinear splitting functions.

In ALARIC, we write the Eikonal factor as
\begin{equation}
    \label{eq:soft_eikonal}
    w_{ik,j}=\frac{W_{ik,j}}{E_j^2}\;,
    \qquad\text{where}\qquad
    W_{ik,j}=\frac{1-\cos\theta_{ik}}{(1-\cos\theta_{ij})(1-\cos\theta_{jk})}\;,
\end{equation}
and perform a partial fraction decomposition of $W_{ik,j}$:
\begin{equation}
    \label{eq:partfrac_soft_matching}
    W_{ik,j}=\bar{W}_{ik,j}^i+\bar{W}_{ki,j}^k\;,
    \qquad\text{where}\qquad
    \bar{W}_{ik,j}^i=\frac{1-\cos\theta_{ik}}{
      (1-\cos\theta_{ij})(2-\cos\theta_{ij}-\cos\theta_{jk})}\;.
\end{equation}
The resulting $\bar{W}_{ik,j}^i$ are strictly positive, thus allowing for an interpretation as the probability of a soft splitting. Furthermore, in contrast to angular ordered showers, our treatment of the Eikonal factor allows to populate the whole phase space including angular correlations, thus also capturing non-global logarithms.

We can now combine the $\bar{W}_{ik,j}^i$ with the regular collinear splitting functions, by subtracting the collinear limit of the Eikonal:
\begin{equation}\label{eq:matching_soft_coll}
    \frac{1}{2p_ip_j}P_{(ij)i}(z)\to \frac{1}{2p_ip_j}P_{(ij)i}(z)
    +\delta_{(ij)i}\,{\bf T}_{i}^2
    \Bigg[\frac{\bar{W}_{ik,j}^i}{E_j^2}-w_{ik,j}^{\rm(coll)}(z)\Bigg]\;,
\end{equation}
which now depend on the direction of the colour spectator $k$.

\begin{figure}[t]
\includegraphics[width=\textwidth]{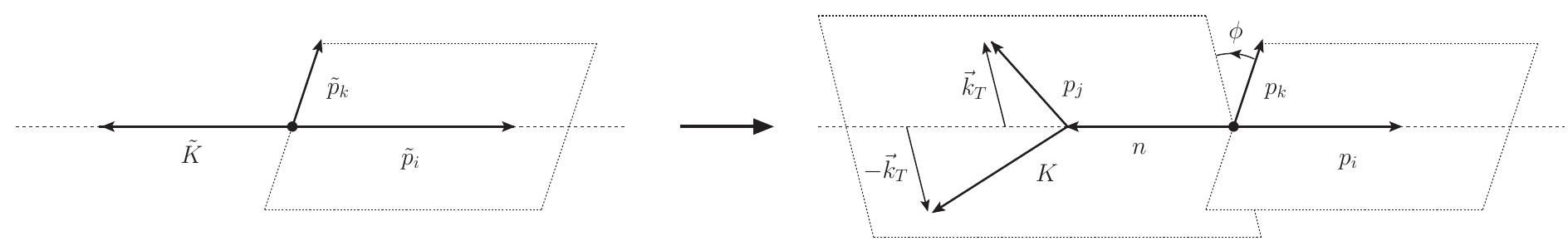}
\caption{The momentum mapping for final-state evolution. After the shift, we boost back from the $K$-frame to the $\tilde{K}$-frame.
\label{fig:kinematics_ff}}
\end{figure}

\paragraph{Recoil scheme}
The second important ingredient is the recoil scheme. Instead of using the colour spectator as recoil partner, we preserve its direction and magnitude, allowing us the define the additional direction entering the collinear splitting functions. Additionally, we preserve the direction of the emitter and compensate the recoil by the sum of all multipole momenta $\tilde{K}$. Furthermore, we require that the invariant mass of the multipole is conserved, leading us to the momenta after the splitting:
\begin{equation}\label{eq:def_pi_K}
    \begin{split}
        p_i&=z\,\tilde{p}_i\;,\\
        p_k&= \tilde{p}_k\;,\\
        p_j&=(1-z)\,\tilde{p}_i+v\big(\tilde{K}-(1-z+2\kappa)\,\tilde{p}_i\big)+k_\perp\;,\\
        K&=\hspace*{11mm}\tilde{K}-v\big(\tilde{K}-(1-z+2\kappa)\,\tilde{p}_i\big)-k_\perp\;,
        \end{split}
\end{equation}
depicted in Fig.~\ref{fig:kinematics_ff}.
Here, $z$ is the momentum fraction of the emitter, $v = p_ip_j/p_i\tilde{K}$ and $\tilde{K}^2/(2\tilde{p}_i\tilde{K})$

Finally, the whole configuration is boosted back into the original frame. Previous emissions are only mildly affected by the boost, which is the basis of the proof of NLL accuracy in Ref.~\cite{Herren:2022jej}. For initial-initial or initial-final dipoles the mapping works similar.

\paragraph{NLO matching}
To combine ALARIC with fixed-order NLO calculations, we follow the MC@NLO method \cite{Frixione:2002ik}. To subtract the double counting, the splitting kernels need to be integrated in $D=4-2\epsilon$ dimensions, which is not easily possible for all shower algorithms. ALARICs similarity to the identified particle subtraction scheme of Ref.~\cite{Catani:1996vz} enables a simple computation of the required terms. We follow Ref.~\cite{Hoche:2018ouj} for the actual calculation. The only non-trivial integral involves the ${\bf H}$-operator \cite{Hoche:2018ouj}:
\begin{equation}\label{eq:h_operator}
\begin{split}
    &\int_0^1\mathrm{d}{z}\, \mathbf{H}_{\tilde{\imath}i}(p_1,\dots,p_i,\dots,p_m;n;z)\\
    &\qquad=- \frac{\alpha_s}{2\pi}\sum_{k=1,k\neq\tilde{\imath}}^m
    \frac{\mathbf{T}_{\tilde{\imath}}\mathbf{T}_{k}}{\mathbf{T}_{\tilde{\imath}}^2}
    \Bigg\{\;\mathcal{K}^{\tilde{\imath}i}
    +\delta_{\tilde{\imath}i}\,\mathrm{Li}_2\bigg(1-
    \frac{2\tilde{p}_i\tilde{p}_k\,\tilde{K}^2}{(\tilde{p}_i\tilde{K})(\tilde{p}_k\tilde{K})}\bigg)
    -\int_0^1\mathrm{d}z\,P^{\tilde{\imath}i}_{\rm reg}(z)
    \ln\frac{(p_ip_k)n^2}{2(p_i n)(p_kn)}\;\Bigg\}\;.
    \end{split}
\end{equation}
The finally integral can not be computed in closed form since $n$, as depicted in Fig.~\ref{fig:kinematics_ff}, depends implicitly on $z$. However, its numerical evaluation is straightforward.
\\
\\
Additional key points such as the choice of evolution variable and the analytic proof of NLL accuracy can be found in Ref.~\cite{Herren:2022jej}.

\section{Results}
\begin{figure}[h]
  \centering
  \begin{minipage}{5.875cm}
    \includegraphics[width=\textwidth]{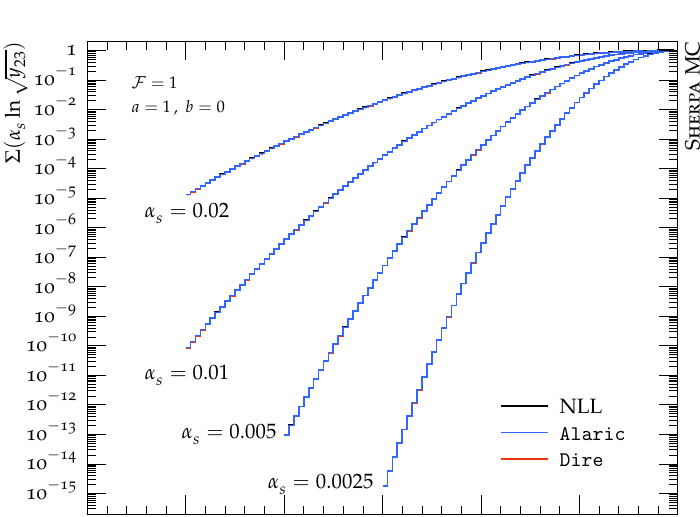}\\[-.8mm]
    \includegraphics[width=\textwidth]{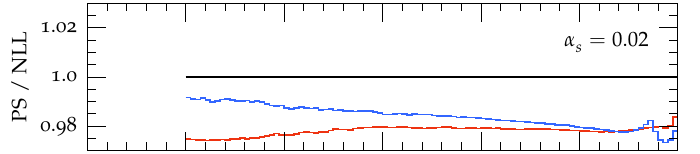}\\[-.8mm]
    \includegraphics[width=\textwidth]{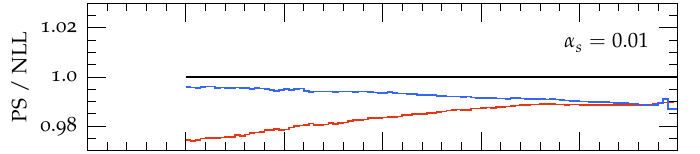}\\[-.8mm]
    \includegraphics[width=\textwidth]{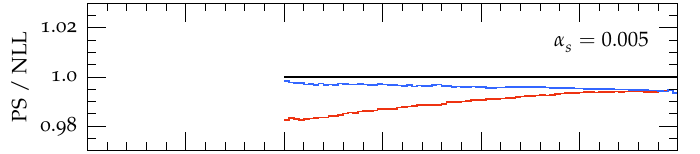}\\[-.8mm]
    \includegraphics[width=\textwidth]{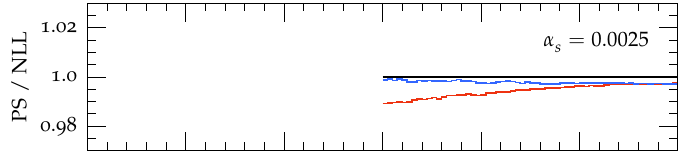}\\[-.8mm]
    \includegraphics[width=\textwidth]{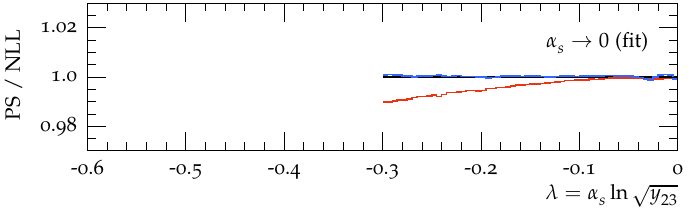}
  \end{minipage}\hfill
  \begin{minipage}{8cm}
  \includegraphics[width=\textwidth]{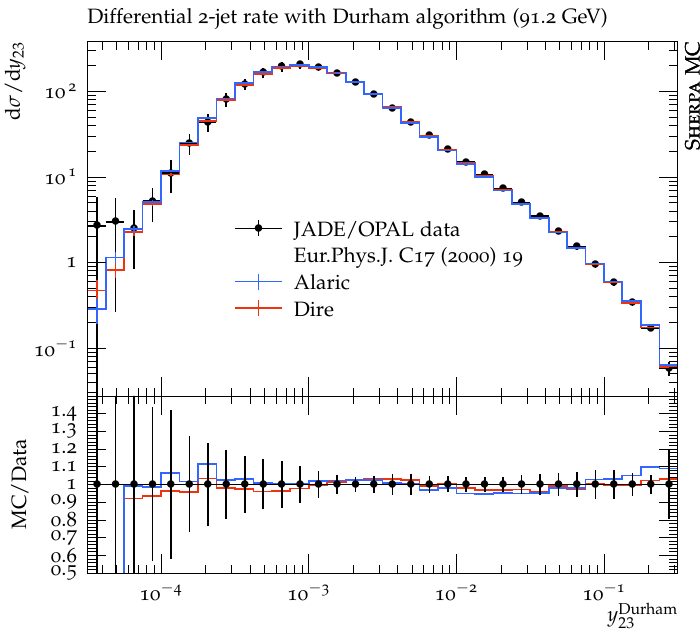}
  \end{minipage}
  \caption{Comparison of the logarithmic accuracy of the leading Lund plane declustering scale in the Cambridge algorithm, $y_{23}$, between ALARIC and DIRE. While ALARIC is NLL accurate for $y_{23}$, DIRE is not. Both algorithms agree well with JADE/OPAL data \cite{JADE:1999zar}.\label{fig:y23}}
\end{figure}

While an analytic proof exists, that the ALARIC recoil scheme is NLL accurate, we can explicitly check it for observables with known NLL resummation results. To this end we tested several observables in $e^+e^-$ collisions following Ref.~\cite{Dasgupta:2020fwr}. In Fig.~\ref{fig:y23} we show the test for the leading Lund plane declustering scale in the Cambridge algorithm, $y_{23}$. While the ALARIC result converges nicely to the NLL result in the limit $\alpha_s \rightarrow 0$, the DIRE algorithm as implemented in SHERPA \cite{Gleisberg:2003xi,Gleisberg:2008ta,Sherpa:2019gpd} does not.

With our preliminary implementation of ALARIC in SHERPA, we can also compare results obtained with ALARIC to $e^+ e^-$ collision data. In Fig.~\ref{fig:y23} we compare to the $y_{23}$ data from JADE/OPAL \cite{JADE:1999zar} and find perfect agreement, despite using a standard hadronization tune. However, differences between ALARIC and DIRE are small, indicating that the logarithmic accurarcy of the shower 

\section{Outlook}
The formulation of ALARIC discussed in these proceedings is sufficient to describe processes in $e^+e^-$ collisions involving only massless quarks. The algorithm has since been extended to properly describe the evolution of massive quarks \cite{Assi:2023rbu} and to proton-proton collisions, including multi-jet merging \cite{Hoche:2024dee}. While some implementation work is still required, ALARIC is on track to be part of an upcoming release of the SHERPA event generator. Consequently, experiments at the LHC will be able to harness fully differential NLL accurate predictions in their analyses.

\section{Acknowledgments}
FH thanks Stefan H\"oche, Frank Krauss, Marek Sch\"onherr and Daniel Reichelt for the fruitful collaboration leading to the ALARIC shower. This research was supported in part by the Swiss National Science Foundation (SNF) under contract 200021-212729. FH acknowledges support by the Alexander von Humboldt Foundation.

\end{document}